\documentclass[final,authoryear,5p,times,twocolumn]{elsarticle}

\usepackage{graphicx}
\usepackage{amssymb}

\usepackage{graphicx}
\usepackage{dcolumn}
\usepackage{bm}
\usepackage{color}
\usepackage{graphicx}
\usepackage{multirow}
\usepackage{tabularx}
\usepackage{amsmath}
\usepackage{booktabs}

\usepackage{array}
\newcommand{\PreserveBackslash}[1]{\let\temp=\\#1\let\\=\temp}
\newcolumntype{C}[1]{>{\PreserveBackslash\centering}p{#1}}
\newcolumntype{R}[1]{>{\PreserveBackslash\raggedleft}p{#1}}
\newcolumntype{L}[1]{>{\PreserveBackslash\raggedright}p{#1}}

\begin{document}

\begin{frontmatter}

\title{Modelling stock correlations with expected returns from investors}

\author[addressa]{Ming-Yuan Yang}
\author[addressb]{Sai-Ping Li}
\author[addressc]{Li-Xin Zhong}
\author[addressa,addressd,addresse]{Fei Ren \corref{cor1}}
\cortext[cor1]{Corresponding author. Address: 130 Meilong Road, P.O. Box 114, School of Business, East China University of Science and Technology, Shanghai 200237, China, Phone: +86 21 64253369.}
\ead{fren@ecust.edu.cn}
\address[addressa]{School of Business, East China University of Science and Technology, Shanghai 200237, China}
\address[addressb]{Institute of Physics, Academia Sinica, Taipei 115 Taiwan}
\address[addressc]{School of Finance, Zhejiang University of Finance and Economics, Hangzhou 310018, China}
\address[addressd]{School of Science, East China University of Science and Technology, Shanghai 200237, China}
\address[addresse]{Research Center for Econophysics, East China University of Science and Technology, Shanghai 200237, China}

\begin{abstract}
Understanding stock correlations is crucial to asset pricing, investor decision-making, and financial risk regulations. However, a microscopic explanation based on agent-based modeling is still lacking. We here propose a model derived from minority game for modeling stock correlations, in which an agent's expected return for one stock is influenced by the historical return of the other stock. Each agent makes a decision based on his expected return with reference to information dissemination and the historical return of the stock. We find that the returns of the stocks are positively (negatively) correlated when agents' expected returns for one stock are positively (negatively) correlated with the historical return of the other. We provide both numerical simulations and analytical studies and give explanations to stock correlations for cases with agents having either homogeneous or heterogeneous expected returns. The result still holds when other factors such as holding decisions and external events are included which broadens the practicability of the model.
%
\end{abstract}


\end{frontmatter}


\section{Introduction}
\label{S1:Introduction}
Stock correlations are high during financial crises, i.e., the stock prices move up or down together. This is a phenomenon in stock markets worldwide, which has drawn the attention of many scholars \citep{Aste-Shaw-Matteo-2010-NJP, Didier-Love-Peria-2012-IJFE, Sharma-Banerjee-2015-PA, Kim-Kim-Lee-2015-PBFJ, Liu-Tse-2012-IJBC}. Understanding this phenomenon is very important for investors and regulators. \cite{Pindyck-Rotemberg-1993-QJE} first found that stock return correlations among US firms are too high to be justified by fundamentals only, while company size and degree of institutional ownership could also influence stock correlations.

Later on, scholars have explained stock correlations from the macroscopic perspective, i.e., the market and industry level. Some studies focus on the correlation of stocks in markets \citep{Morck-Wu-Yeung-2000-JFE, Didier-Love-Peria-2012-IJFE, Dang-Moshirian-2015-JFE, Li-Peng-2017-EMd}. \cite{Morck-Wu-Yeung-2000-JFE} and \cite{Dang-Moshirian-2015-JFE} found that stock correlations tended to be higher in poor and emerging markets with poor institutions and property rights. \cite{Didier-Love-Peria-2012-IJFE} found that stock price co-movements were driven largely by financial linkages between markets. \cite{Li-Peng-2017-EMd} noted that economic policy uncertainty were able to affect the comvoements of both Chinese and US stocks. Some studies focus on the correlation of stocks in industries \citep{Kallberg-Pasquariello-2008-JEF, Anton-polk-2014-JF, Huang-An-Fang-Gao-Wang-Sun-2016-RSER, Peng-Zhu-Jia-2017-EMd}. \cite{Kallberg-Pasquariello-2008-JEF}, \cite{Anton-polk-2014-JF} showed that firms in the same industry had correlated earnings and therefore returns. \cite{Huang-An-Fang-Gao-Wang-Sun-2016-RSER} showed that the information dissemination through firms in the same industry would cause a stock price co-movement. \cite{Peng-Zhu-Jia-2017-EMd} adduced a strong evidence of size effect on stock price synchronicity in Chinese oil firms.

Other studies have explained stock correlations from the perspective of firms. Several studies have shown that a firm's specific information is the primary cause of stock correlations \citep{Hameed-Morck-Yeung-2015-RFS, Liu-Wu-Li-Li-2015-ESA}. Others have explained stock correlations with complex network methods \citep{Qiao-Xia-Li-2016-PloS, Zhang-Li-Ye-Li-Nagi-2015_IEEEIS, Li-Tian-Ma-Li-2012-cnJMSC}. \cite{Qiao-Xia-Li-2016-PloS} have found that stock network linkages could explain stock correlations. \cite{Zhang-Li-Ye-Li-Nagi-2015_IEEEIS} simulated information dissemination on a firm network, and predicted the stock price co-movements with the ECM (Energy Cascading Model). \cite{Li-Tian-Ma-Li-2012-cnJMSC} have found that the prices of stocks moved together along with the shareholder link network and the director network. All these studies suggest that the more closely firms are linked in the network, the higher correlations they have.

Recently, scholars have also studied stock correlations from the perspective of investors' actions \citep{Li-Tang-Liu-2011-cnMW, Eun-Wang-Xiao-2015-JFE, Li-Zhao-2016-NAJEF,  Frijns-Verschoor-Zwinkels-2017-JIFMIM, Filzen-Schutte-2017-NAJEF, Veldkamp-2006-REStu, Mondria-2010-JET}. Some empirical studies show that investors' sentiment and irrational actions will cause stock correlations \citep{Li-Tang-Liu-2011-cnMW, Eun-Wang-Xiao-2015-JFE, Li-Zhao-2016-NAJEF, Frijns-Verschoor-Zwinkels-2017-JIFMIM}. \cite{Li-Tang-Liu-2011-cnMW} found that investors' irrational actions could cause stock price co-movements by studying the Chinese A-share markets. {\cite{Eun-Wang-Xiao-2015-JFE} found that culture influences could cause stock price synchronicity by affecting investors' trading activities and a country's information environment. \cite{Li-Zhao-2016-NAJEF} found that investors preferred to trade the stocks of firms which stayed in their own city, and this behavior could cause stock price co-movements. \cite{Frijns-Verschoor-Zwinkels-2017-JIFMIM} found that stock return comvemnents were driven by the non-fundamental part of return, which was mainly the investor sentiment. Other studies focus on information dissemination through investors \citep{Filzen-Schutte-2017-NAJEF,Mondria-2010-JET,Veldkamp-2006-REStu}. \cite{Filzen-Schutte-2017-NAJEF} found that stock price would co-move when investors could acquire low cost information about other firms by empirical analysis. \cite{Mondria-2010-JET} proved that asset prices would co-move when agents could simultaneously get the price information of the both assets to forecast asset prices through analytical analysis. \cite{Veldkamp-2006-REStu} proposed a theoretical model to demonstrate that when rational agents could get information of stocks simultaneously, stock prices would co-move with no correlation between stock fundamentals. To date, only few studies are devoted to explain the microcosmic reason of stock correlations from the perspective of agents' actions, most of which are based on theoretical analysis. Our study is motivated by research on stock correlations with agent-based models and reveal its microcosmic mechanism by modeling the trading behavior of individual agents.

Minority game is a highly successful agent-based model that can be used to explain most stylized facts in financial markets \citep{Challet-Marsili-Zhang-2001a-PA}. The standard minority game was first proposed by \cite{Challet-Zhang-1997-PA} based on the El Farol bar problem \citep{Arthur-1994-AER}. It describes a system in which heterogeneous agents adaptively compete for scarce resource, and it captures some key features of a generic market mechanism and the basic interactions between agents and public information. As a successful agent-based model to simulate the stock market, minority game has been under rapid development and largely used in memory size \citep{Johnson-Hui-Zheng-Hart-1999-JPA}, evolution mechanism \citep{Challet-Marsili-Zhang-2001-QF}, strategy selection \citep{DHulst-Rodgers-1999-PA, Zhong-Zheng-Zheng-Hui-2005-PRE}, stock market simulations \citep{Yeung-Wong-Zhang-2008-PRE}, market impact \citep{Barato-Mastomatteo-Bardoscia-Marsili-2013-QF} and agent behavior \citep{Zhang-Huang-Wu-Su-Lai-2016-SR}. One of the most important applications of the minority game is its modeling for multi-assets. \cite{Bianconi-DeMartino-Ferreira-Marsili-2008-QF} first studied the minority game in which agents might invest two assets to investigate speculative trading effects in stock markets. \cite{Martino-Castillo-Sherrington-2007-JSM} introduced a multi-assets minority game to solve the strategy frequency problem. Inspired by their works, we here introduce a new model based on minority game in which agents may trade two stocks, and focus on its explanation to stock correlations. This extension makes the model closer to the real stock markets, and can be widely applied to asset
pricing, investor decision-making and financial risk regulations.

We propose that an agent's expected return on one stock is influenced by its own historical return and the historical return of the other stock. This expected return is modeled with reference to studies on the theory of information dissemination and the facts that exist in real stock markets. Some studies on information dissemination theory imply that stock price movements are influenced by the spread of information among investors \citep{Bordino-Battiston-Caldarelli-Cristelli-Ukkonen-Weber-2012-PloS, Heiberger-2015-PloS}. In addition, \cite{Zhang-Li-Ye-Li-Nagi-2015_IEEEIS} and \cite{Chuang-2016-NAJEF} found that the stock price co-movement is due to the dissemination of information between firms, and \cite{Chen-Tan-Zheng-2015-SR} proposed that for the firms shared common information in the same industry their stocks had the same price movement. \cite{Marcet-2017-EMR} also proposed that when information dissemination through analysts, there would be price co-movement across Latin American stocks. These studies demonstrated that information from different firms could be obtained and spread by investors, and the spread of information could lead to the changes of stock prices which in turn could affect the price movements of other stocks. In a real stock market, we can see that investors trade stocks by referring to the performance of other stocks in the same industry. Sometimes they even refer to the performance of other stocks in the competitive industry, and trade according to the opposite sign of the price changes of relevant stocks. These investors' behaviors will ultimately have an impact on the stock correlations. The information dissemination theory proposed in literature and the above  phenomena exist in real stock markets provide the theoretical and realistic basis for the establishment of the expected return in our model.

We also propose that an agent makes a decision based on his or her expected return and historical returns of the stock in the minority game. The simulation results of our model suggest that the stock returns are positively (negatively) correlated when the expected returns are positively (negatively) correlated, and the correlation of stock returns is proportional to the correlation of the expected returns of the stock. Compared to previous studies, this paper makes two contributions. First, we propose a new model based on the MG model to explain the correlations between stocks from the perspective of agent-based modeling. Second, we theoretically demonstrate that investors' expectation of one stock influenced by another stock essentially leads to the correlation between the two by both numerical simulation and analytical analysis.

The remainder of this paper is organized as follows. We present our model and basic assumptions in section 2. The results and analysis of the simulation are given in section 3. Finally, we conclude in section 4.

\section{Model and assumptions}
\label{S2:Model and assumptions}

To study the correlation between individual stocks, we assume the simple case when there are only two stocks traded by agents. Our model takes the form of a repeated game with an odd number $N$ of agents who must choose an independent decision of buying or selling actions according to their strategies. A strategy is a way of decision-making in which an investor makes a decision according to the information available, including the historic returns in last $m$ time steps and the expected return for a specific stock. Since there are $2^{m}$ possible bit strings of historical returns, two possible options (up or down) for each agent's expectation, and two decisions of buying or selling actions, there are a total of $2^{2^{m+1}}$ strategies. Table \ref{TB:1} illustrates an agent's strategy for the transaction of stock $j$ in our model.

\setlength\tabcolsep{7pt}
\begin{table}
\small
\begin{center}
\begin{minipage}{88mm}
\caption{\label{tabone}A strategy of agent $i$ for stock $j$. Here $r_j (t-m)$ $(j=1, 2)$ denotes the historical return of the stock $j$ $(j=1, 2)$ at time $t-m$, and $r_{j, i}^e (t)$ $(j=1, 2)$ denotes the expected return of agent $i$ for stock $j$ $(j=1, 2)$ at time $t$. $+$ $(-)$ means the return is positive (negative). $\sigma_j^i (t)$ $(j=1, 2)$ denotes the decision that agent $i$ $(i=1,..., N)$ makes for stock $j$ $(j=1, 2)$ according to the history and expectation. $1$ $(-1)$ means buying (selling) 1 unit of stock.}
{\begin{tabular}{@{}ccccccr}\toprule
  & History&&& Expectation && Decision\\
  \cline{1-3}\cline{5-5}\cline{7-7}
$r_j(t-m)$& ...& $r_j(t-1)$&& $r^e_{j, i}(t)$&& $\sigma^i_j(t)$\\
\toprule
  $+$& ...& $+$&& $+$&& 1\\
  $-$& ...& $-$&& $-$&& -1\\
  $...$& ...& $...$&& $...$&& ...\\
  $+$& ...& $+$&& $-$&& 1\\
  $+$& ...& $-$&& $-$&& -1\\
\toprule
\end{tabular}}
\label{TB:1}
\end{minipage}
\end{center}
\end{table}

As seen from Table \ref{TB:1}, the history of a stock is the stock return series in the last $m$ steps, which can be accessed by all agents. Expectation in the table is an agent's expected return, which is determined by the agent himself/herself. The agent employs a specific strategy to determine the trading decision according to the current history and the expected return. The decision is denoted by 1 and -1 respectively in the table, representing buying or selling 1 unit of stock. Since the agents do not have cash in our simplified model, which makes it difficult to calculate the trading capacity of each agent, we assume that each agent has a trading capacity of one share of each stock in each transaction. We also consider the case when the agent can take a holding position, making a decision not buying or selling, which is a common practice to study incentives of the agents to participate in the market \citep{Yeung-Wong-Zhang-2008-PRE, Challet-Marsili-Zhang-2001-QF, Giardina-Bouchaud-2003-EPJB}. As one will see in the results, the holding decision does not change the main results of our model.

The expected return is defined as follows. The expected return of agent $i$ for stock 1 is
\begin{equation}
r^e_{1,i}(t)=a_1r_1(t-1)+b_{1,i}r_2(t-1),
 \label{Eq:1}
\end{equation}
and the expected return of agent $i$ for stock 2 is
\begin{equation}
r^e_{2,i}(t)=a_2r_2(t-1)+b_{2,i}r_1(t-1),
\label{Eq:2}
\end{equation}
where  $r_{j, i}^e (t)$ $(j=1, 2)$ is the expected return of agent $i$ for stock $j$ $(j=1, 2)$ at time $t$, and $r_j (t-1)$ $(j=1, 2)$ is the return of stock $j$ $(j=1, 2)$ at time $t-1$. $a_j$ $(j=1, 2)$ is the first-order autocorrelation coefficient of the stock $j$ and $b_{j, i}$ $(j=1, 2)$ denotes the impact of the first-order lag return of the other stock on stock $j$ for agent $i$.

We propose the expected return with reference to other studies on the information dissemination theory. \cite{Zhang-Li-Ye-Li-Nagi-2015_IEEEIS} and \cite{Chuang-2016-NAJEF} both demonstrated that the dissemination of information between firms caused stock price co-movement by constructing the network of different firms. \cite{Chen-Tan-Zheng-2015-SR} proposed that stocks in the same industry shared common characteristics, which resulted in the synchronous prices rises and falls.
Since stock price return is one kind of the most important information that diffuses in stock markets and studied by scholars, we here introduce the stock price return. \cite{Mondria-2010-JET} found that changes in one asset affected both asset prices and might lead to asset price co-movement when investors could choose linear combinations of asset payoffs to update information about the assets. Based on these studies, we propose that the expected return of agents to be in the form of a linear combination of stock returns in our model.

We can also see that investors trade stocks by considering the performance of other stocks in real stock markets, and this provides realistic basis for the assumption of the above expectation. The underlying reason for the investors' behaviors that focus on the performances of multi assets may lie in the fundamental correlations between these stocks, which have business relationships between each other. For example, investors will refer to the performance of other stocks in the same industry when they trade stocks. In particular, investors would tend to pay attention on the performance of the leading stock, especially when the leading stock has just experienced a boom or crash. Sometimes they even refer to the performance of stocks not in the same industry but in the same industry chain. These behaviors will ultimately affect investors' expectations.

To begin with, each agent $i$ randomly picks $S$ strategies from the full strategy space, sticks with them throughout the game, and keeps track of the cumulative performance of his or her trading strategy $s$ $(s=1,...,S)$ by assigning a score $U_j^{i, s}(t)$ $(j=1,2$; $t=0,1,..., T)$ to each of them. The initial scores of the strategies are set to be zero. Initially, agent $i$ randomly selects a strategy $s_i$ among his/her $S$ own strategies, and the trading decision is also randomly selected since there are no historical returns and expected return at $t=0$.
At time step $t>0$, each agent $i$ adopts a strategy $s_i$ with the highest score. If $U_j^{i, s_i}(t)$ is the highest at time $t$, the decision of agent $i$ to trade stock $j$ is $\sigma_j^{i}(t)=\sigma_j^{i, s_i}(t)$ corresponding to the current historical returns and the expected return of stock $j$.

After all agents have made their decisions of actions, the excess demand of stock $j$ at time $t$ is calculated as
\begin{equation}
A_j(t)=\sum_{i=1}^N\sigma^i_j(t).
\label{Eq:3}
\end{equation}
$A_j(t)$ measures the imbalance between buyers (demand) and sellers (supply), which represents the force of driving the price up and down respectively, and is commonly used to update the price of the stock \citep{Yeung-Wong-Zhang-2008-PRE, Challet-Marsili-Zhang-2001-QF, Challet-Marsili-Zhang-2000-PA}.

The price of stock $j$ at time $t$ is updated according to
\begin{equation}
P_j(t)=P_j(t-1)+sgn[A_j(t)]|A_j(t)|^{0.5},
\label{Eq:4}
\end{equation}
where the square root in the formula is commonly used in the price dynamics of the MG model \citep{Yeung-Wong-Zhang-2008-PRE, Challet-Marsili-Zhang-2000-PA}, which is also supported by the evidence of empirical studies \citep{Zhang-1999-PA}. The return of stock $j$ at time $t$ is
\begin{equation}
r_j(t)=log(P_j(t))-log(P_j(t-1)).
\label{Eq:5}
\end{equation}

The score of strategy $s$ $(s=1,...,S)$ holding by the agent $i$ is then updated as below
\begin{equation}
U_j^{i, s}(t)=U_j^{i, s}(t-1)+g_j^{i, s}(t),
\label{Eq:6}
\end{equation}
where the payoffs $g_j^{i, s}(t)$ to strategy $s$ is
\begin{equation}
g_j^{i, s}(t)=-sgn[A_j(t)]\sigma_j^{i, s}(t).
\label{Eq:7}
\end{equation}
The payoff is calculated with the excess demand, which is in line with the price update. Suppose the excess demand is positive (negative) and the decision of the strategy is to sell (buy), the agent sells (buys) a unit of stock with a transaction price at $t$, higher (lower) than the current price at $t-1$. Therefore, the strategies which help the agents buy stocks at cheaper prices and sell stocks at higher price are rewarded, and vice versa. The rule that the agents in the minority group would win is the so-called minority game model.

The agent then repeatedly chooses the best strategy from his/her fixed $S$ strategies according to their updated scores, and makes the decision of buying or selling action according to the newly updated historical returns and expected return. The expected return of agent $i$ for stock $j$ is updated according to Eqs.(\ref{Eq:1}) and (\ref{Eq:2}). The stock price and score of the strategy are subsequently updated following Eqs.(\ref{Eq:4}) and (\ref{Eq:6}). The model evolves by repeating the steps listed above.

\section{Simulation results and analysis}

For simplicity and the computational efficiency of the model, we choose the case where $N=1001$, $m$ and $S$ are relatively small e.g., $m = 1$ and $S = 2$. We also perform a simulation for $m=2$ and $S=2$, and the simulation result is similar to that for $m=1$ and $S=2$. The simulation result becomes stable when $T\sim1000$. Therefore, we choose $T$ to be 1000. The initial price of our model is set to be 2000, large enough to ensure that the price is positive throughout the entire evolution period, and the results are robust to different initial prices.

We also need to determine the values of $a_j$ $(j=1, 2)$ and $b_{j, i}$ $(j=1, 2;$ $i=1,..., N)$. We know that $a_j $ is the first-order autocorrelation coefficient of the stock $j$, and $b_{j, i}$ is the impact that the first-period lag return of the other stock on stock $j$ for agent $i$. Thus, we set the values of $a_j$ $(j=1, 2)$ based on the first-order autoregressive coefficient of the stocks in the real market. We study 15 A-share stocks trade on the Shanghai and Shenzhen Stock Exchanges from January 4, 2011 to December 31, 2015. The data source is from Beijing Gildata RESSET Data Technology Co., Ltd, see http://www.resset.cn/. The calculation results are shown in Table \ref{TB:2}.

\setlength\tabcolsep{2.6pt}
\begin{table}
\small
\centering
\begin{minipage}{88mm}
\caption{First-order autoregressive coefficients of 15 A-share stocks trade on Shanghai and Shenzhen Stock Exchanges from January 4, 2011 to December 31, 2015.}
{\begin{tabular}{@{}crrrrr}\toprule
{Stock Code}&{2011}&{2011-2012}&{2011-2013}&{2011-2014}&{2011-2015}\\
\toprule
000001&-0.1129&-0.0728&0.0057&-0.0066&0.0161\\
000002&-0.0824&-0.0563&0.0413&0.0142&0.0430\\
000004&0.1659&0.1067&0.0642&0.0919&0.1398\\
000005&0.0330&0.0217&0.0080&0.0658&0.2845\\
000006&-0.0819&-0.0412&-0.0526&-0.0444&0.0907\\
000007&0.0948&0.1181&0.1057&0.1388&0.1359\\
000008&0.1396&0.1834&0.1511&0.2418&0.2124\\
000009&0.0815&0.0475&0.0171&0.0544&0.0788\\
600000&-0.0593&-0.0424&0.0106&-0.0032&0.0032\\
600004&0.0529&0.0083&-0.0255&0.0268	&0.0544\\
600005&0.0561&-0.0029&-0.0075&-0.0092&0.0818\\
600006&0.0452&0.0428&0.0346&0.0991&0.1285\\
600007&-0.0983&-0.0942&-0.0559&-0.0364&0.0647\\
600008&0.0589&0.0299&0.0745&0.0815&0.1354\\
600009&-0.1029&-0.0810&-0.0449&-0.0110&0.0140\\
\toprule
\end{tabular}}
\label{TB:2}
\end{minipage}
\end{table}

As seen in Table \ref{TB:2}, when the time window increases to five years, the first-order autoregressive coefficients of all stocks are greater than 0. Therefore, we take the value of $a_j$ $(j=1, 2)$ to be between 0 and 1 in the model. Indeed, we have verified that the first-order autoregressive coefficients of the simulated data under the restriction $a_j\in (0, 1]$ are consistent with those of the empirical data. We consider nine combinations of different values of $a_j$, namely $a_1=\{0.1, 0.5, 1\}$ and $a_2=\{0.1, 0.5, 1\}$ in our simulation. Given the values of $a_j$, we now set the values of $b_{j, i}$ $(j=1, 2;$ $i=1,..., N)$. We will set the values of $b_{j, i}$ in two cases: all agents with the same expectation and agents with different expectations. When all agents have the same expected return, we have $b_{j, i}=b_j$ $(i=1,..., N)$. We here assume that the values of $b_{j, i}$ $(j=1,2)$ are between -1 and 1, and the interval is 0.1. The cases in which $b_{j, i}$ $(j=1, 2)$ are greater than 1 and less than -1 will be presented in subsection~\ref{sec:3.2}.

After deciding the value for each parameter, one could now obtain the return series of stock $j$ $(j=1, 2)$. To discuss the correlation between two stocks, we calculate the Pearson's correlation coefficient. The Pearson's correlation coefficient between two stock return series $r_1(t)$ and $r_2(t)$ is defined as
\begin{equation}
\rho_r=\frac{Cov(r_1, r_2)}{\sqrt{Var(r_1)}\sqrt{Var(r_2)}},
\label{Eq:8}
\end{equation}
where $Cov(r_1, r_2)$ is the covariance of the two stock return series, and $Var(r_1)$ and $Var(r_2)$ are the variances of the two stock return series respectively. We perform 50 runs and take the mean correlation coefficient as the final simulation result.

\subsection{Simulation results and analysis for all agents with the same expectation}
\label{sec:3.1}
\subsubsection{Simulation results}
\label{sec:3.1.1}

When all agents have the same expected return, we have $b_{j, i}=b_j$ $(j=1,2;$ $i=1,..., N)$. We here assume that the values of $b_j$ $(j=1,2)$ are between -1 and 1, and the interval is 0.1. The simulation results for $a_1=a_2$ and ${a_1}\neq{a_2}$ are shown in Figs. \ref{Fig:1} and \ref{Fig:2} respectively.

\begin{figure}
\begin{center}
\begin{minipage}{88mm}
\centering\includegraphics[width=3.5in]{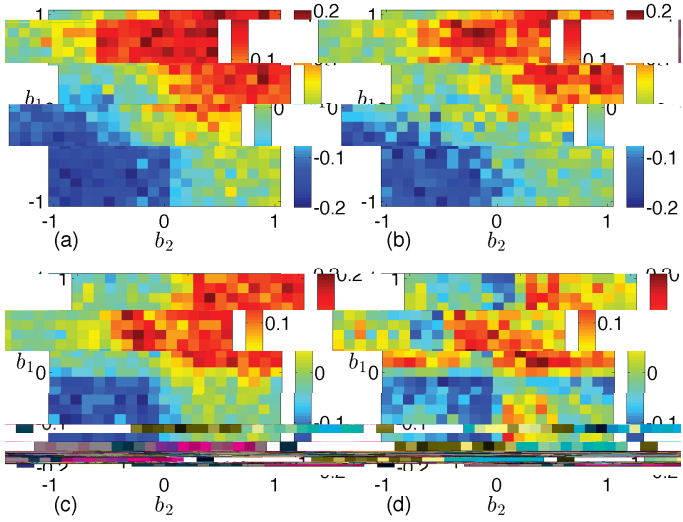}
\caption{The correlation coefficients between the stock returns for: (a) $m=2$, $S=2$ and $a_1=a_2=1$, (b) $m=1$, $S=2$ and $a_1=a_2=1$, (c) $m=1$, $S=2$ and $a_1=a_2=0.5$, (d) $m=1$, $S=2$ and $a_1=a_2=0.1$. }
\label{Fig:1}
\end{minipage}
\end{center}
\end{figure}

From the simulation result in Fig. \ref{Fig:1}(a), we can see that the correlation coefficients between the stock returns are positive (negative) when $b_1>0$ and $b_2>0$ ($b_1<0$ and $b_2<0$), and the absolute values of the correlation coefficients are large when $b_1$ and $b_2$ are near 1 or -1. We also see that the simulation result for $m=1$ and $S=2$ in Fig. \ref{Fig:1}(b) is not much different from the result for $m=2$ and $S=2$ in Fig. \ref{Fig:1}(a). Hence, we only analyze the results for $m=1$ and $S=2$ in the following. In Fig. \ref{Fig:1}(c), the correlation coefficients between the stock returns are positive (negative) when $b_1>0$ and $b_2>0$ ($b_1<0$ and $b_2<0$), and the absolute values of the correlation coefficients are large when $b_1$ and $b_2$ are near 0.5 or -0.5.  In Fig. \ref{Fig:1}(d), the correlation coefficients between the stock returns are small in most cases, and the absolute values of the correlation coefficients are large when $b_1$ and $b_2$ are near 0.1 or -0.1.

\begin{figure}
\begin{center}
\begin{minipage}{88mm}
\centering\includegraphics[width=3.5in]{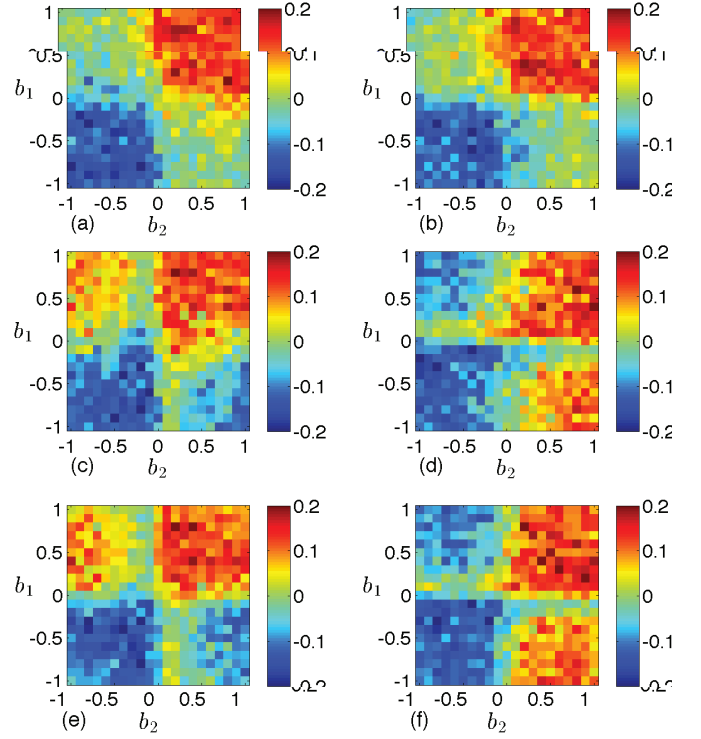}
\caption{The correlation coefficients between the stock returns with parameters $m=1$ and $S=2$ for: (a) $a_1=1$ and $a_2=0.5$, (b) $a_1=0.5$ and $a_2=1$, (c) $a_1=1$ and $a_2=0.1$, (d) $a_1=0.1$ and $a_2=1$, (e) $a_1=0.5$ and $a_2=0.1$, (f) $a_1=0.1$ and $a_2=0.5$. }
\label{Fig:2}
\end{minipage}
\end{center}
\end{figure}

From the simulation results in Fig. \ref{Fig:2}, for example, for $a_1=1$ and $a_2=0.1$ in Fig. \ref{Fig:2}(c), it can be seen that the correlation coefficients between the stock returns are positive (negative) when $b_1>0$ $(b_1<0)$, and the absolute value of the correlation coefficients are small when $b_2$ is near 0. When $b_1>0$ $(b_1<0)$, the expected return of stock 2 is almost only affected by stock 1 and the expected return of stock 1 is affected by both stocks, hence the correlation coefficients between two stock returns are positive (negative). In addition, the expected return of stock 2 is almost unaffected by either stocks when $b_2$ is near 0, hence the correlation coefficient between the stock returns is very small. Moreover, the results in Figs. \ref{Fig:2}(e) and \ref{Fig:2}(f) are similar to the  results in Figs. \ref{Fig:2}(c) and \ref{Fig:2}(d) respectively, while the results in Figs. \ref{Fig:2}(a) and \ref{Fig:2}(b) are similar to results in Figs. \ref{Fig:1}(b) and \ref{Fig:1}(c) respectively. Therefore, we will take the simulation results in Fig. \ref{Fig:1}(b) as a representative to analyze.

We have also checked the results for the case that the agent can take a holding position, i.e., the agents make trading decisions $\sigma_j^{i}(t)=1,-1,$ or 0, which represent buying or selling a unit of stock, or doing nothing. Here, we show the simulation results for $m=1$, $S=2$ and $a_1=a_2=1$ in Fig. \ref{Fig:3}. In comparison with the results for the model without holding position shown in Fig. \ref{Fig:1}(b), the correlation of the present model becomes relatively weaker, but the basic behavior remains the same. The introduction of the holding position does not change the main results, hence this feature will not be included in our model.

\begin{figure}
\begin{center}
\begin{minipage}{88mm}
\centering\includegraphics[width=3.5in]{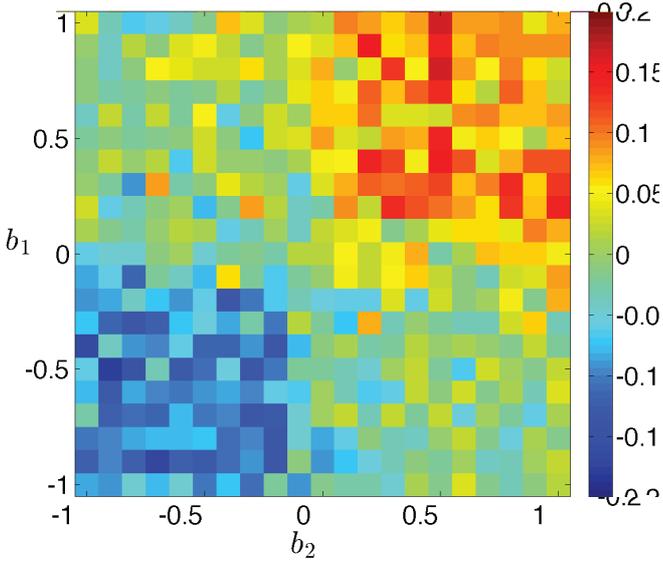}
\caption{The correlation coefficients between the stock returns for the case that the agent can take a holing position with parameters $m=1$, $S=2$ and $a_1=a_2=1$.}
\label{Fig:3}
\end{minipage}
\end{center}
\end{figure}

It is worth to study how much investors' expectations of stock returns account for the stock return correlation when other factors are included in their trading decisions. For instance the exogenous factor introduced in \citet{Papadopoulos-Coolen-2008-JPA}, which considers a case when the market is impacted by external events that may represent the actions of market regulators, or other major natural or political events. We consider the impacts of the external events on the stock returns, upon which the investors make decisions to buy or sell, therefore study their influences on investors' trading decisions. In doing so, the internal and external contributions to the overall excess demand are introduced as
\begin{equation}
A_j(t)=A^{int}_j(t)+ A^{ext}_j(t),
\label{Eq:9}
\end{equation}
where $A^{int}_j(t)$ is the excess demand defined in Eq. (\ref{Eq:3}), and $A^{ext}_j(t)$ represents the contribution of external events, which has a general formula
\begin{equation}
A^{ext}_j(t)=(-1)^{\theta (t)}\tilde{A_j}E_j(t).
\label{Eq:10}
\end{equation}
To make $A^{ext}_j(t)$ more realistic, we use the empirical data of the news of A-share stocks trade on Shanghai Stock Exchange during the period form December 16, 2013 to November 22, 2016 to calculate the average probability of the occurrence of an external event per minute, and use it to determine the probability $p(E_j(t)=1)=0.0082$. The basic results remain to be the same if $p(E_j(t)=1)$ is not too large. $\theta (t)$ is a randomly generated integer, which determines the sign of the impact, positive or negative when $\theta (t)$ is even or odd. $\tilde{A_j}$ reflects the impact strength of the external event, which is measured in units of the standard deviation $s_j$ of $A^{int}_j(t)$, i.e., $\tilde{A_j}=ks_j$. With the new overall excess demand defined above, the price is therefore updated according to Eq. (\ref{Eq:4}), and the return is consequently generated according to Eq. (\ref{Eq:5}).

The results of the correlation coefficients between the stock returns for the case when  external events are included in investors' trading decisions for $k=1,2,3,4$ are shown in Fig. \ref{Fig:5}. As the impact strength of external events increases, the correlation between the stocks becomes relatively smaller. However, the main results are essentially similar to the ones shown in Fig. \ref{Fig:1}(b). The returns of the stocks are still correlated under the interference of exogenous factor like external events, and this shows that the stock return correlations are largely determined by the mechanism of the expected returns proposed in our model.

\begin{figure}
\begin{center}
\begin{minipage}{88mm}
\centering\includegraphics[width=3.5in]{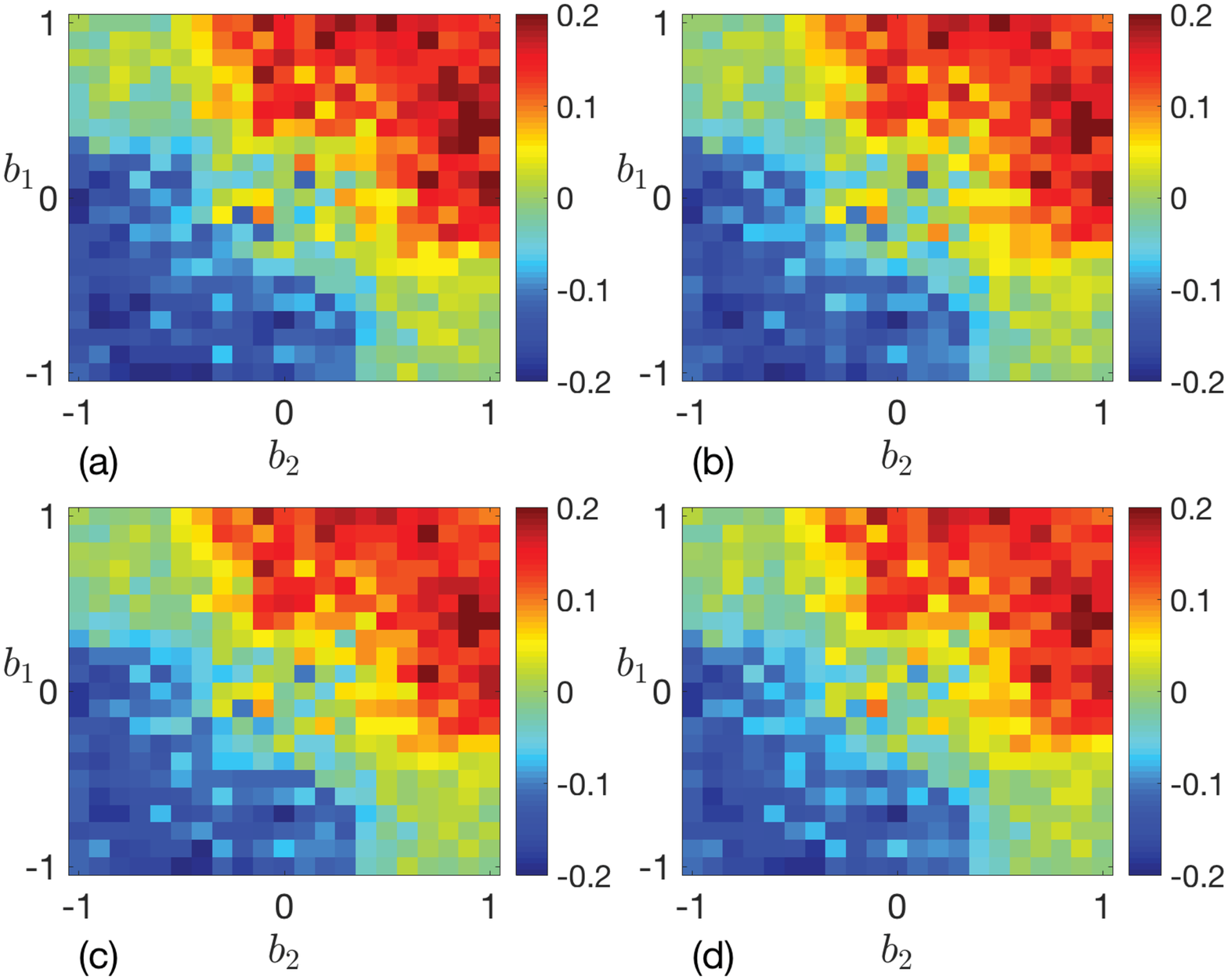}
\caption{The correlation coefficients between the stock returns for the case that external events are included into investors' trading decisions with parameters $m=1$, $S=2$ and $a_1=a_2=1$ for: (a) $k=1$, (b) $k=2$, (c) $k=3$, (d) $k=4$.}
\label{Fig:5}
\end{minipage}
\end{center}
\end{figure}

\subsubsection{Analysis of results}
\label{sec:3.1.2}

From the simulation results, regularities in the correlation coefficient between the stock returns can be seen. However, what causes the regularities, and why the correlation coefficient is very close to 0 in some cases, while in other cases it is greater than zero and less than zero? To explain these phenomena, it is necessary to analyze the source of the stock correlation. From the assumptions of the model, we know that the correlation between the stock returns is mainly determined by the agent's expected returns of the stocks. Since the expected return of the agent on one stock takes into account the influence of the other stock, the correlation between the expected returns is bound to affect the correlation between the stock returns. To verify our argument, we should analyze the relationship between the stock return and its expected return. First, we draw the scatter plots of both stocks. We choose the simulation results when $m=1$, $S=2$ and $a_1=a_2=1$ for the scatter plots, which is shown in Fig. \ref{Fig:6}.

\begin{figure}
\begin{center}
\begin{minipage}{88mm}
\centering\includegraphics[width=3.5in]{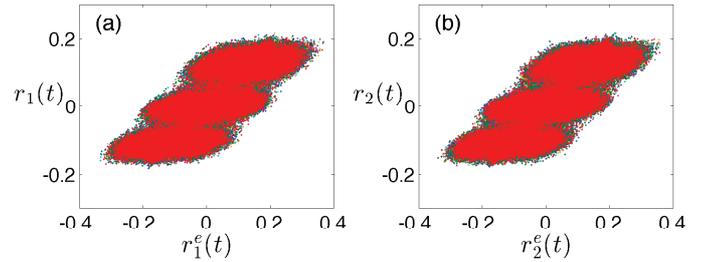}
\caption{Scatter plot of the return versus its expected return for (a) stock 1 and (b) stock 2 with parameters $m=1$, $S=2$ and $a_1=a_2=1$. Colors correspond to the returns of different runs.}
\label{Fig:6}
\end{minipage}
\end{center}
\end{figure}

It can be seen that the expected returns and returns of the stocks cluster in three regions in the scatter plots. This is mainly due to the fact that the memory size $m$ is small. In a minority game, one will display a multi-peak distribution for the return when the memory size is small. It can also be seen that there is a linear relationship between the return and its expected return. We therefore propose a linear regression model between the return and its expected return as follows

\begin{equation}
r_j(t)=\beta_0+{\beta_1}{r^e_j}(t)+\epsilon,
\label{Eq:11}
\end{equation}
where $r_j$ $(j=1, 2)$ and $r_j^e$ $(j=1, 2)$ represent the return and the expected return of stock $j$ respectively. The regression analysis of the stocks are done separately and the results are given in Table \ref{TB:3}.

\setlength\tabcolsep{7.5pt}
\begin{table}
\small
\begin{center}
\begin{minipage}{88mm}
\caption{\label{tabone}Regression results of the stocks for $a_1=a_2=1$.}
{\begin{tabular}{@{}ccccccccc}\toprule
&{Stock 1}&&&&{Stock 2}\\
\cline{1-3}\cline{5-7}
$\beta_1$&p-value&$R^2$&&$\beta_1$&p-value&$R^2$\\
\cline{1-3}\cline{5-7}
0.7174&0& 0.7485&&0.7172&0& 0.7476\\
\toprule
\end{tabular}}
\label{TB:3}
\end{minipage}
\end{center}
\end{table}

As shown in Table \ref{TB:3}, there is a significant positive correlation between the return and its expected return. Although regression results suggest that there is a significant correlation between the two, just how the correlation between the two stock expected returns affects the correlation between the two stock returns cannot be understood from the regression model. We therefore study the correlation between the two stock expected returns in order to explain the correlation between two stock returns.

From Figs. \ref{Fig:1} and \ref{Fig:2}, we know that the correlation between the stock returns is mainly determined by the values of $b_j$ $(j=1, 2)$. We therefore concentrate on how the values of $b_j$ affect the correlation between the expected returns of the stocks, and thus the correlation between the stock returns. We here put forth some concepts for the following discussion. $\Delta{r_j}(t-1)=r_j(t-1)-r_j(t-2), (j=1, 2)$ is the change of stock $j$'s return defined by the first-order difference at time $t-1$. $\Delta{r_j^e}(t)$ $(j=1, 2)$ is the change of the expected return of stock $j$ defined by the first-order difference at time $t$. From Eqs. (\ref{Eq:1}) and (\ref{Eq:2}), the changes of the stock expected returns at time $t$ are:
\begin{equation}
\left\{
             \begin{array}{lr}
             \Delta{r_1^e}(t)={a_1}\Delta{r_1}(t-1)+{b_1}\Delta{r_2}(t-1) &  \\
             \Delta{r_2^e}(t)={a_2}\Delta{r_2}(t-1)+{b_2}\Delta{r_1}(t-1). &
             \end{array}
\right.
\label{Eq:12}
\end{equation}
When $a_1=a_2=1$, we have
\begin{equation}
\left\{
             \begin{array}{lr}
             \Delta{r_1^e}(t)=\Delta{r_1}(t-1)+{b_1}\Delta{r_2}(t-1)  &  \\
             \Delta{r_2^e}(t)=\Delta{r_2}(t-1)+{b_2}\Delta{r_1}(t-1).  &
             \end{array}
\right.
\label{Eq:13}
\end{equation}
Four separate cases \uppercase\expandafter{\romannumeral1}-\uppercase\expandafter{\romannumeral4} will be considered for different values of $b_j$ $(j=1, 2)$ when $a_1=a_2=1$.

\uppercase\expandafter{\romannumeral1}. $0<b_1<1$ and $0<b_2<1$

(a) If $\Delta{r_1}(t-1)>0$ and $\Delta{r_2}(t-1)>0$, then
\begin{equation}
\left\{
             \begin{array}{lr}
             \Delta{r_1^e}(t)=\Delta{r_1}(t-1)+{b_1}\Delta{r_2}(t-1)>0 &  \\
             \Delta{r_2^e}(t)=\Delta{r_2}(t-1)+{b_2}\Delta{r_1}(t-1)>0. &
             \end{array}
\right.
\label{Eq:14}
\end{equation}
Here, the expected returns of both stocks increase at the same time, and the correlation between the expected returns is positive.

(b) If $\Delta{r_1}(t-1)<0$ and $ \Delta{r_2}(t-1)<0$, the analysis is similar to \uppercase\expandafter{\romannumeral1} (a). Here, the expected returns of both stocks decrease at the same time, and the correlation between the expected returns is positive.

(c) If $\Delta{r_1}(t-1)>0$ and $\Delta{r_2}(t-1)<0$, we discuss the possibilities for four cases of combinations of $\Delta{r_1^e(t)}$ and $\Delta{r_2^e}(t)$.

According to Eq. (\ref{Eq:13}), the condition satisfying $\Delta{r_1^e}(t)<0$ and $\Delta{r_2^e}(t)>0$ is
\begin{equation}
         -\frac{\Delta{r_2}(t-1)}{b_2}<\Delta{r_1}(t-1)<-{b_1}\Delta{r_2}(t-1).
                   \label{Eq:15A}
\end{equation}
Since $-{\Delta{r_2}(t-1)}/{b_2}>-{b_1}\Delta{r_2}(t-1)$ when $\Delta{r_2}(t-1)<0$, Eq. (\ref{Eq:15A}) cannot be satisfied, therefore the case when $\Delta{r_1^e}(t)<0$ and $\Delta{r_2^e}(t)>0$ is impossible.

For the case when $\Delta{r_1^e}(t)>0$ and $\Delta{r_2^e}(t)<0$ are satisfied, according to Eq. (\ref{Eq:13}), the condition will be
\begin{equation}
         -{b_1}\Delta{r_2}(t-1)<\Delta{r_1}(t-1)<-\frac{\Delta{r_2}(t-1)}{b_2}.
                   \label{Eq:15B}
\end{equation}
The value range of $\Delta{r_1}(t-1)$ becomes smaller when $b_1$ and $b_2$ approaches 1, and the probability of satisfying this condition becomes smaller. Therefore, the negative correlation between the expected returns becomes weaker.

We next discuss the other two cases that the expected returns are positively correlated.
If $\Delta{r_1^e}(t)>0$ and $ \Delta{r_2^e}(t)>0$ are satisfied, according to Eq. (\ref{Eq:13}), the condition will be
\begin{equation}
          \Delta{r_1}(t-1)>-\frac{\Delta{r_2}(t-1)}{b_2}.
          \label{Eq:15C}
\end{equation}
When $b_2$ approaches 1, the lower limit value of $\Delta{r_1}(t-1)$ approaches $-\Delta{r_2}(t-1)$. Therefore, the value range of $\Delta{r_1}(t-1)$ becomes larger, and the probability of satisfying $\Delta{r_1^e}(t)>0$ and $\Delta{r_2^e}(t)>0$ also becomes larger. The probability that the expected returns of both stocks increase simultaneously is larger, i.e., the positive correlation between the expected returns becomes stronger.
Similarly, if $\Delta{r_1^e}(t)<0$ and $\Delta{r_2^e}(t)<0$ are satisfied, the positive correlation of the expected returns is stronger when $b_1$ approaches 1.

Among the four cases above, three of them are valid, which include two cases when the expected returns are positively correlated with large probability. Therefore the correlation are positive on average, and the positive correlation becomes stronger when $b_j$ $(j=1, 2)$ approaches 1.

(d) If $\Delta{r_1}(t-1)<0$ and $\Delta{r_2}(t-1)>0$, the analysis is similar to \uppercase\expandafter{\romannumeral1} (c), see Appendix A \uppercase\expandafter{\romannumeral1} (d) for details. Here, the positive correlation between the expected returns becomes stronger when $b_j$ $(j=1, 2)$ approaches 1.

From \uppercase\expandafter{\romannumeral1} (a)-\uppercase\expandafter{\romannumeral1} (d), we see that the expected returns of the stocks are positively correlated, and the expected return correlation is stronger when $b_j$ $(j=1, 2)$ is closer to 1.

\uppercase\expandafter{\romannumeral2}. $-1<b_1<0$ and $-1<b_2<0$

The analysis is similar to \uppercase\expandafter{\romannumeral1}, see Appendix A \uppercase\expandafter{\romannumeral2} for details. Here, the expected returns of the stocks are negatively correlated, and the expected return correlation is stronger when $b_j$ $(j=1, 2)$ is closer to -1.

\uppercase\expandafter{\romannumeral3}. $0<b_1<1$ and $ -1<b_2<0$

(a) If $\Delta{r_1}(t-1)>0$ and $ \Delta{r_2}(t-1)>0$, $\Delta{r_1^e}(t)$ is always positive. There are two possible cases for $\Delta{r_2^e}(t)$, namely  $\Delta{r_2^e}(t)>0$ and $\Delta{r_2^e}(t)<0$.

If $\Delta{r_1^e}(t)>0$ and $\Delta{r_2^e}(t)>0$ are satisfied, according to Eq. (\ref{Eq:13}), the condition will be
\begin{equation}
          \Delta{r_2}(t-1)>-{b_2}\Delta{r_1}(t-1).
          \label{Eq:16}
\end{equation}
When $b_2$ approaches -1, the lower limit value of $\Delta{r_2}(t-1)$ approaches $\Delta{r_1}(t-1)$. Therefore, the value range of $\Delta{r_2}(t-1)$ becomes smaller, and the probability of satisfying $\Delta{r_1^e}(t)>0$ and $\Delta{r_2^e}(t)>0$ becomes smaller. The probability that the expected returns of both stocks increase simultaneously becomes smaller, i.e., the positive correlation between the expected returns becomes weaker.

Similarly, if $\Delta{r_1^e}(t)>0$ and $\Delta{r_2^e}(t)<0$ are satisfied, the negative correlation between the expected returns becomes stronger when $b_2$ approaches -1.

(b) If $\Delta{r_1}(t-1)<0$ and $\Delta{r_2}(t-1)<0$, the analysis is similar to \uppercase\expandafter{\romannumeral3} (a), see Appendix A \uppercase\expandafter{\romannumeral3} (b) for details. Here, the positive (negative) correlation between the expected returns becomes weaker (stronger) when $b_2$ approaches -1.

(c) If $\Delta{r_1}(t-1)>0$ and $\Delta{r_2}(t-1)<0$, $\Delta{r_2^e}(t)$ is always negative. There are two cases for $\Delta{r_1^e}(t)$, namely  $\Delta{r_1^e}(t)<0$ and $\Delta{r_1^e}(t)>0$.

If $\Delta{r_1^e}(t)<0$ and $\Delta{r_2^e}(t)<0$ are satisfied, according to Eq. (\ref{Eq:13}), the condition will be
\begin{equation}
           \Delta{r_2}(t-1)<-\frac{\Delta{r_1}(t-1)}{b_1}.
           \label{Eq:17}
\end{equation} 
When $b_1$ approaches 1, the upper limit value of $\Delta{r_2}(t-1)$ approaches $-\Delta{r_1}(t-1)$. Therefore, the value range of $\Delta{r_2}(t-1)$ becomes larger, and the probability of satisfying $\Delta{r_1^e}(t)<0$ and $\Delta{r_2^e}(t)<0$ becomes larger. The probability that the expected returns of both stocks decrease simultaneously becomes larger, i.e.,the positive correlation between the stock expected returns becomes stronger.

Similarly, if $\Delta{r_1^e}(t)>0$ and $\Delta{r_2^e}(t)<0$ are satisfied, the negative correlation between the stock expected returns is weaker when $b_1$ approaches 1.

(d) If $\Delta{r_1}(t-1)<0$ and $\Delta{r_2}(t-1)>0$, the analysis is similar to \uppercase\expandafter{\romannumeral3} (c), see Appendix A \uppercase\expandafter{\romannumeral3} (d) for details. Here, the negative (positive) correlation between the stock expected returns becomes weaker (stronger) when $b_1$ approaches 1.

From \uppercase\expandafter{\romannumeral3} (a)-\uppercase\expandafter{\romannumeral3} (d), we observe that the positive correlation between the expected returns is stronger when $b_1$ is closer to 1. We also observe that the negative correlation between the stock expected returns is stronger when $b_2$ is closer to -1.

\uppercase\expandafter{\romannumeral4}. $-1<b_1<0$ and $0<b_2<1$

The analysis is similar to \uppercase\expandafter{\romannumeral3}, see Appendix A \uppercase\expandafter{\romannumeral4} for details. The positive correlation between the stock expected returns is stronger when $b_2$ is closer to 1. We also observe that the negative correlation between the expected returns is stronger when $b_1$ is closer to -1.

From Fig. \ref{Fig:6} and Table \ref{TB:3}, we can see that there is a significant positive correlation between the returns and expected returns. We summarize the results of simulated returns based on the correlation of expected returns in the following.

\begin{enumerate}
\item The correlation between the stock returns is positive when $0<b_1<1$ and $0<b_2<1$, and gets stronger as $b_j$ $(j=1, 2)$ increases.

\item The correlation between the stock returns is negative when $-1<b_1<0$ and $-1<b_2<0$, and the correlation is stronger when $b_j$ $(j=1, 2)$ decreases.

\item The correlation between the stock returns is weak when $0<b_1<1$ and $-1<b_2<0$, and it changes from negative to positive as $b_j$ $(j=1, 2)$ increases.

\item The correlation between the stock returns is weak when $-1<b_1<0$ and $0<b_2<1$, and it changes from negative to positive as $b_j$ $(j=1, 2)$ increases.
\end{enumerate}
We can also observe the correlation between the stock returns from Fig. \ref{Fig:1}(b). The results in Fig. \ref{Fig:1}(b) and the analytic results in the above thus agree with each other.

\subsection{Simulation results and analysis for agents with different expectations}
\label{sec:3.2}

In a real stock market, agents are heterogeneous and their expected returns are not exactly the same. Indeed, we can consider the heterogeneous agents that have different values for both $a_j$ and $b_{j, i}$ $(j=1, 2;$ $i=1,..., N)$, according to the definition of expected return in Eqs. (\ref{Eq:1}) and (\ref{Eq:2}). However, the parameter $b_{j, i}$ can reflect the information dissemination through stocks, which plays a core role in explaining the correlation between stocks. Besides, we have also studied the case with a uniform distribution of $a_j \sim U(0,1)$, and found that the results remain similarly. Hence, we mainly focus on the parameter $b_{j, i}$ and show the simulation results for its different values in the following discussion. We will make an additional assumption that each agent $i$ in the model matches a unique $b_{j,i}$ $(j=1, 2)$, which is subject to a uniform distribution, i.e., $b_{j,i}\sim U(c_j-\delta_j, c_j+\delta_j)$. The center of the distribution is $c_j$, and the distribution range is $2\delta_j$.

It is worth to note that the results do not depend on the specific formula of the distribution, but only the symmetry feature of the distribution significantly affect the results. As one will see in the following discussions on the distributions with varying centers and ranges, the correlation between the stocks is positive or negative when the distribution center is bias to positive or negative, which may be driven by the forces of news or big events in real stock markets. For the distribution which is symmetric about 0, things cancel out each other and the stocks therefore become uncorrelated.

\subsubsection{Simulation results for the distribution with varying distribution center and fixed distribution range}
\label{sec:3.2.1}

The simulation result for $a_1=a_2=1$ and $\delta_1=\delta_2=1$ is shown in Fig. \ref{Fig:7}. Centers $c_1$ and $ c_2$ are from -1 to 1, and the intervals are 0.2.
The result in Fig. \ref{Fig:7} suggests that the correlation coefficient is proportional to $c_1$ and $ c_2$.

\begin{figure}[htb]
\begin{center}
\begin{minipage}{88mm}
\centering\includegraphics[width=3.5in]{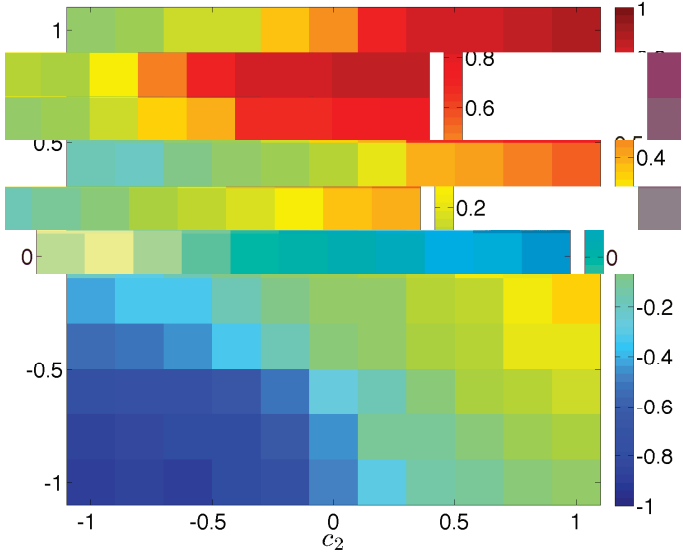}
\caption{The correlation coefficients between the stock returns when $m=1$, $S=2$, $a_1=a_2=1$ and $\delta_1=\delta_2=1$.}
\label{Fig:7}
\end{minipage}
\end{center}
\end{figure}

\subsubsection{Analysis of results for the distribution with varying distribution center and fixed distribution range}
\label{sec:3.2.2}

Similar to the analysis in subsubsection~\ref{sec:3.1.2}, we first analyze the relationship between the return and its expected return. We draw the scatter plots of the returns versus expected returns of the stocks for four cases of $c_1$ and $c_2$ in Fig. \ref{Fig:8}, namely, \uppercase\expandafter{\romannumeral1}. $0<c_1<1$ and $0<c_2<1$, \uppercase\expandafter{\romannumeral2}. $-1<c_1<0$ and $-1<c_2<0$, \uppercase\expandafter{\romannumeral3}. $0<c_1<1$ and $-1<c_2<0$, \uppercase\expandafter{\romannumeral4}. $-1<c_1<0$ and $0<c_2<1$. A linear relationship between the return and its expected return can be observed in Fig. \ref{Fig:8}.

\begin{figure}[htb]
\begin{center}
\begin{minipage}{88mm}
\centering\includegraphics[width=3.5in]{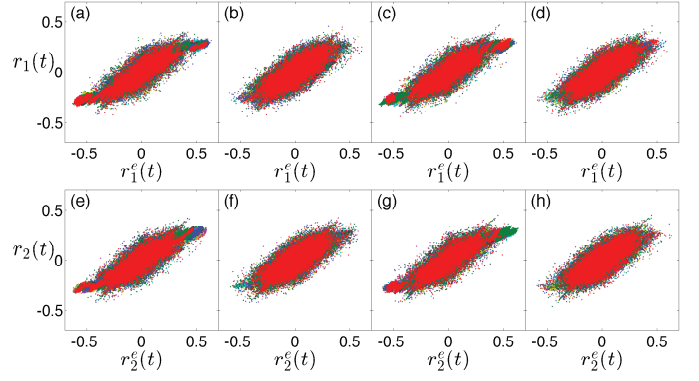}
\caption{ Scatter plots of the return versus its expected return for stock 1 with different values of $c_1$ and $c_2$: (a) $0<c_1<1$ and $0<c_2<1$, (b) $-1<c_1<0$ and $-1<c_2<0$, (c) $0<c_1<1$ and $-1<c_2<0$, (d) $-1<c_1<0$ and $0<c_2<1$. Scatter plots of the returns versus expected returns for stock 2 with different values of $c_1$ and $c_2$: (a) $0<c_1<1$ and $0<c_2<1$, (b) $-1<c_1<0$ and $-1<c_2<0$, (c) $0<c_1<1$ and $-1<c_2<0$, (d) $-1<c_1<0$ and $0<c_2<1$. Colors correspond to the returns of different runs, with parameters $m=1$, $S=2$ and $a_1=a_2=1$.}
\label{Fig:8}
\end{minipage}
\end{center}
\end{figure}

\setlength\tabcolsep{2pt}
\begin{table}[htb]
\footnotesize
\centering 
\caption{Regression results of two stocks for $a_1=a_2=1$ and $\delta_1=\delta_2=1$.}
{\begin{tabular}{rrcccccccc}\toprule
\multicolumn{1}{c}{\multirow {2}{*}{$c_1$}}&\multicolumn{1}{c}{\multirow {2}{*}{$c_2$}} &&{Stock 1}&&&&{Stock 2}\\
 \cline{3-5}\cline{7-9}
&&$\beta_1$&p-value&$R^2$&&$\beta_1$&p-value&$R^2$\\
\cline{1-9}

$ 0<c_1<1$ &$ 0<c_2<1$&0.6195&0&0.8806&&0.6266&0&0.8847\\
$ 0<c_1<1$ &$-1<c_2<0$&0.7219&0&0.7335&&0.7106&0&0.7387\\
$-1<c_1<0$ &$-1<c_2<0$&0.6249&0&0.8844&&0.6193&0&0.8849\\
$-1<c_1<0$ &$ 0<c_2<1$&0.7173&0&0.7473&&0.7173&0&0.7371\\
\toprule
\end{tabular}}
\label{TB:4}
\end{table}

We therefore perform a regression analysis for the stocks using Eq. (\ref{Eq:11}), and the results are presented in Table \ref{TB:4}. One can see from Table \ref{TB:4} that there is a significant positive correlation between the return and its expected return when $b_1$ and $b_2$ are uniformly distributed.

Using the linear relationship between the return and its expected return, one can obtain the correlation between the returns from the correlation between the expected returns. From Eq. (\ref{Eq:12}), the variations of the expected returns of stock $j$ $(j=1, 2)$ for agent $i$ $(i=1, ..., N)$ at time $t$ are
\begin{equation}
\left\{
 \begin{array}{lr}
\Delta r^e_{1, i}(t) =a_1\Delta r_1(t-1)+b_{1, i}\Delta r_2(t-1) &\\
\Delta r^e_{2, i}(t)=a_2\Delta r_2(t-1)+b_{2,i}\Delta{r_1}(t-1). &
\end{array}
\right.
\label{Eq:18}
\end{equation}
The average variations of the expected returns for stock 1 and stock 2 are
\begin{equation}
\left\{
             \begin{array}{lr}
             \overline{\Delta{r_1^e}}(t)=\frac{1}{N}\sum\limits_{i=1}^Na_1\Delta r_1(t-1)+\frac{1}{N}\sum\limits_{i=1}^Nb_{1,i}\Delta r_2(t-1)   &  \\
             \overline{\Delta{r_2^e}}(t)=\frac{1}{N}\sum\limits_{i=1}^Na_2\Delta r_2(t-1)+\frac{1}{N}\sum\limits_{i=1}^Nb_{2,i}\Delta r_1(t-1),   &
             \end{array}
\right.
\label{Eq:19}
\end{equation}
where $b_{1,i}$ and $b_{2,i}$ are the values of $b_1$ and $b_2$ for agent $i$ respectively. Since $b_1$ and $b_2$ are subject to the uniform distribution, we have
\begin{equation}
\frac{1}{N}\sum_{i=1}^Nb_{1,i}=c_1,   \   \   \   \    \frac{1}{N}\sum_{i=1}^Nb_{2,i}=c_2.
\label{Eq:20}
\end{equation}

From Eqs. (\ref{Eq:19}) and (\ref{Eq:20}), we see that the average variations of the expected returns for the stocks when $a_1=a_2=1$ are

\begin{equation}
\left\{
             \begin{array}{lr}
             \overline{\Delta{r_1^e}}(t)=\Delta r_1(t-1)+c_1\Delta r_2(t-1)   &  \\
             \overline{\Delta{r_2^e}}(t)=\Delta r_2(t-1)+c_2\Delta r_1(t-1).  &
             \end{array}
\right.
\label{Eq:21}
\end{equation}

We now study the correlation of the expected returns from Eq. (\ref{Eq:21}) for different values of $c_1$ and $c_2$. Eq. (\ref{Eq:21}) here takes the same form as Eq. (\ref{Eq:13}), with $c_1$ replacing $b_1$, and $c_2$ replacing $b_2$. Similar to the analysis for fixed values of $b_1$ and $b_2$, four separate cases will be considered here.

\uppercase\expandafter{\romannumeral1}. $0<c_1<1$ and $0<c_2<1$

The positive correlation between the expected returns is stronger when $c_1$ and $c_2$ are close to 1, and is weaker when $c_1$ and $c_2$ are close to 0.

\uppercase\expandafter{\romannumeral2}. $-1<c_1<0$ and $-1<c_2<0$

The negative correlation between the expected returns is stronger when $c_1$ and $c_2$ are close to -1, and is weaker when $c_1$ and $c_2$ are close to 0.

\uppercase\expandafter{\romannumeral3}. $0<c_1<1$ and $-1<c_2<0$

The positive correlation between the expected returns becomes stronger when $c_1$ approaches 1 and $c_2$ approaches 0. The negative correlation between the expected returns becomes stronger when $c_1$ approaches 0 and $c_2$ approaches -1.

\uppercase\expandafter{\romannumeral4}. $-1<c_1<0$ and $0<c_2<1$

The positive correlation between the expected returns becomes stronger when $c_1$ approaches 0 and $c_2$ approaches 1. The negative correlation between the expected returns becomes stronger when $c_1$ approaches -1 and $c_2$ approaches 0.

From Fig. \ref{Fig:8} and Table \ref{TB:4}, we see that there is a significant positive correlation between the return and its expected return. Therefore, we can obtain the correlation between the returns from the analytic results of \uppercase\expandafter{\romannumeral1}-\uppercase\expandafter{\romannumeral4}, which is summarized below.

\begin{enumerate}
\item The correlation of the returns is positive when $0<c_1<1$ and $0<c_2<1$, and becomes stronger as $c_j$ $(j=1, 2)$ increases.

\item The correlation of the returns is negative when $-1<c_1<0,$ and $-1<c_2<0$, and becomes stronger as $c_j$ $(j=1, 2)$ decreases.

\item The correlation of the returns is weak when $0<c_1<1$ and $-1<c_2<0$, and changes from negative to positive as $c_j$ $(j=1, 2)$ increases.

\item The correlation of the returns is weak when $-1<c_1<0$ and $0<c_2<1$, and changes from negative to positive as $c_j$ $(j=1, 2)$ increases.
\end{enumerate}

We can also observe the correlation between the stock returns from Fig. \ref{Fig:7}, which agrees with the analysis in this subsection.

\subsubsection{Simulation results for the distribution with fixed distribution center and varying distribution range}
\label{sec:3.2.3}

The simulation results for $a_1=a_2=1$ and $c_1=c_2=0$ are shown in Fig. \ref{Fig:9}. The ranges of $\delta_1$ and $\delta_2$ are from 1 to 5, and the intervals are 0.5.
From Fig. \ref{Fig:9}, we can see that the correlation coefficients are small, and the correlation between the stock returns is weak.

\begin{figure}[htb]
\begin{center}
\begin{minipage}{88mm}
\centering\includegraphics[width=3.5in]{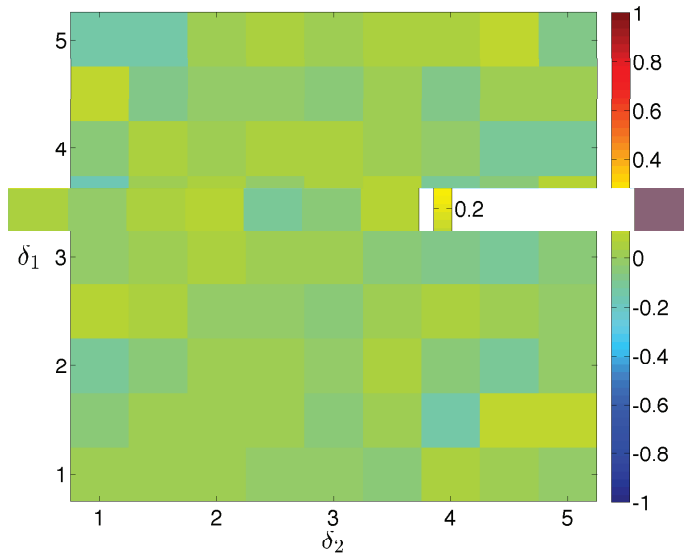}
\caption{ The correlation coefficients between the stock returns for $m=1$, $S=2$, $a_1=a_2=1$ and $c_1=c_2=0$.}
\label{Fig:9}
\end{minipage}
\end{center}
\end{figure}

\begin{figure}[htb]
\begin{center}
\begin{minipage}{88mm}
\centering\includegraphics[width=3.5in]{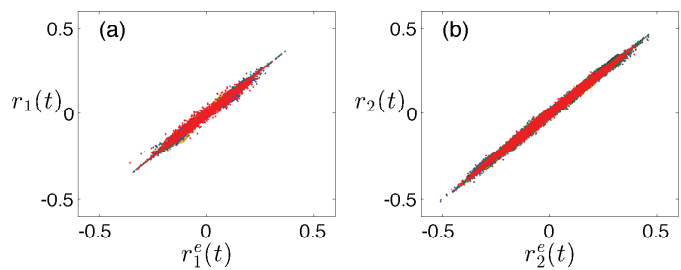}
\caption{Scatter plot of the return versus its expected return for (a) stock 1 and (b) stock 2 with parameters $m=1$, $S=2$ and $a_1=a_2=1$. Colors correspond to the returns of different runs.}
\label{Fig:10}
\end{minipage}
\end{center}
\end{figure}

\subsubsection{Analysis of results for the distribution with fixed distribution center and varying distribution range}
\label{sec:3.2.4}

Scatter plots of the returns versus expected returns of the stocks when $c_1=c_2=0$ are shown in Fig. \ref{Fig:10}. A linear relationship between the return and its expected return can be observed in the figure. We therefore carry out a regression analysis of the stocks separately with Eq. (\ref{Eq:11}), and the results are given in Table \ref{TB:5}.

\setlength\tabcolsep{7.5pt}
\begin{table}
\small
\begin{center}
\begin{minipage}{88mm}
\caption{Regression results of  stocks 1 and 2 for $a_1=a_2=1$ and $c_1=c_2=0$.}
{\begin{tabular}{ccccccccc}\toprule
 &{Stock 1}&&&&{Stock 2}\\
\cline{1-3}\cline{5-7}
$\beta_1$&p-value&$R^2$&&$\beta_1$&p-value&$R^2$\\
\cline{1-3}\cline{5-7}
0.9991&0& 0.9918&&1.0003&0& 0.9985\\
\toprule
\end{tabular}}
\label{TB:5}
\end{minipage}
\end{center}
\end{table}

From the regression results when $b_1$ and $b_2$ are uniformly distributed, the regression of both stocks passed the significance test. There is a significant correlation between the return and its expected return, and the correlation is positive.

From Eq. (\ref{Eq:19}), we get
\begin{equation}
\left\{
             \begin{array}{lr}
             \overline{\Delta{r_1^e}}(t)=a_1\Delta r_1(t-1)+c_1\Delta r_2(t-1)   &  \\
             \overline{\Delta{r_2^e}}(t)=a_2\Delta r_2(t-1)+c_2\Delta r_1(t-1).  &
             \end{array}
\right.
\label{Eq:22}
\end{equation}
When $a_1=a_2=1$ and $ c_1=c_2=0$, we have
\begin{equation}
\left\{
             \begin{array}{lr}
             \overline{\Delta{r_1^e}}(t)=\Delta r_1(t-1)   &  \\
             \overline{\Delta{r_2^e}}(t)=\Delta r_2(t-1).  &
             \end{array}
\right.
\label{Eq:23}
\end{equation}

In this case, we can easily see that the mean value of the expected return is only relevant to the stock return itself. Hence, the expected returns of the stocks are uncorrelated. From Fig. \ref{Fig:10} and Table \ref{TB:5}, we see that the return and its expected return are strongly correlated, implying that stock returns are not correlated. This is confirmed by the simulation results in Fig. \ref{Fig:9}.
From the simulation results and the analysis of the model, we find that stock correlations are proportional to the values of $b_1$ and $b_2$, whether they are fixed or subject to uniform distributions. We also find that the expected returns of agents have a direct impact on the stock correlations.

\section{Conclusion}
\label{sec:4}
In this paper, we study stock correlations by using a model based on minority game, in which we propose that an agent's expected return of one stock is influenced by the historical return of the other stock, and an agent makes a decision with his or her expected return and the historical return of the stock. We model the stock correlation for all agents with the same expectation, and find that the investor's expectation regarding stock return is an important factor for the stock correlation. We also find that stock returns are positively (negatively) correlated when the expectation of returns are positively (negatively) correlated, and the correlation of stock returns is proportional to the correlation of stock expected returns. We then model the stock correlation for agents with different expectations, and the simulation results are similar to the results when all agents have the same expectation. The simulation results also suggest that the center of the distribution has a significant influence on the stock correlation but the range of the distribution has no influence on the stock correlation.
These results remain to be true when we consider other factors such as external events in financial markets. We have therefore demonstrated the underlying mechanism that investors' expectation return of one stock is influenced by another stock would lead to a correlation between two stocks in our model. Two stocks have strong (weak) correlations when investors believe that the stocks have strong (weak) impacts on each other.

To the best of our knowledge, this is the first time that one models the correlations between stocks from the perspective of agent-based modeling. Our model is derived from the standard MG model, in which the agents are allowed to trade multi-assets simultaneously. The expected returns of the agents are modeled with reference to information dissemination in financial markets. To improve the practicability of the model, we further introduce variables that model external news and events in financial markets and fix their values from the analysis of real data. These features make our model simulate real stock markets more closely, and help to expand its practical implications in many issues dealing with multi-assets or systematic problems. The model not only can reveal the microscopic mechanism underlying the stock correlations, but also can be applied to asset pricing, investor decision-making and financial risk regulations.

\section{Acknowledgments}
This work was partially supported by the National Natural Science Foundation (Nos. 10905023, 71131007 and 71371165), the Humanities and Social Sciences Fund sponsored by Ministry of Education of the People's Republic of China (No. 17YJAZH067), the Collegial Laboratory Project of Zhejiang Province (No. YB201628), and the Fundamental Research Funds for the Central Universities (2015). We would also like to thank the anonymous reviewers for their valuable comments and suggestions which help to improve the present work.

\section*{Appendix A}

In this appendix, we provide more detailed descriptions of the analysis presented in subsection~\ref{sec:3.1.2}.

\uppercase\expandafter{\romannumeral1}. $0<b_1<1$ and $ 0<b_2<1$

(d) If $\Delta{r_1}(t-1)<0$ and $\Delta{r_2}(t-1)>0$, we discuss the possibilities for four cases of combinations of $\Delta{r_1^e(t)}$ and $\Delta{r_2^e}(t)$.

According to Eq. (\ref{Eq:13}), the condition satisfying $\Delta{r_1^e}(t)>0$ and $\Delta{r_2^e}(t)<0$ is
 \begin{equation}
                 -{b_1}\Delta{r_2}(t-1)<\Delta{r_1}(t-1)<-\frac{\Delta{r_2}(t-1)}{b_2}.
                   \label{Eq:A1}
\end{equation}
Since $-{\Delta{r_2}(t-1)}/{b_2}<-{b_1}\Delta{r_2}(t-1)$ when $\Delta{r_2}(t-1)>0$, Eq. (\ref{Eq:A1}) cannot be satisfied, therefore this case is impossible.

For the case when $\Delta{r_1^e}(t)<0$ and $\Delta{r_2^e}(t)>0$ are satisfied, according to Eq. (\ref{Eq:13}), the condition will be
\begin{equation}
    -\frac{\Delta{r_2}(t-1)}{b_2}<\Delta{r_1}(t-1)<-{b_1}\Delta{r_2}(t-1).
     \label{Eq:A2}
\end{equation}
Under this condition, the expected returns are negatively correlated with a smaller probability as $b_j$ $(j=1, 2)$ is closer to 1.

We next discuss the other two cases when the expected returns are positively correlated. If $\Delta{r_1^e}(t)>0$ and $\Delta{r_2^e}(t)>0$ are satisfied, according to Eq. (\ref{Eq:13}), the condition will be
\begin{equation}
           \Delta{r_1}(t-1)>-{b_1}\Delta{r_2}(t-1).
           \label{Eq:A3}
\end{equation}
When $b_1$ approaches 1, the lower limit value of $\Delta{r_1}(t-1)$ approaches $-\Delta{r_2}(t-1)$. Therefore, the value range of $\Delta{r_1}(t-1)$ becomes larger, and the probability of satisfying $\Delta{r_1^e}(t)>0$ and $ \Delta{r_2^e}(t)>0$ also becomes larger. The probability that the expected returns of both stocks increase simultaneously is larger, i.e., the positive correlation of the expected returns becomes stronger.
Similarly, if $\Delta{r_1^e}(t)<0$ and $\Delta{r_2^e}(t)<0$ are satisfied, the positive correlation of the expected returns becomes stronger when $b_2$ approaches 1.

For two of the three valid cases, the expected returns are positively correlated with large probability. Therefore, the correlation are positive on average, and the positive correlation becomes stronger when $b_j$ $(j=1, 2)$ approaches 1.

\uppercase\expandafter{\romannumeral2}. $-1<b_1<0$ and $-1<b_2<0$

(a) If $\Delta{r_1}(t-1)>0$ and $\Delta{r_2}(t-1)>0$, we discuss the possibilities for four cases of combinations of $\Delta{r_1^e(t)}$ and $\Delta{r_2^e}(t)$.

According to Eq. (\ref{Eq:13}), the condition satisfying $\Delta{r_1^e}(t)<0$ and $\Delta{r_2^e}(t)<0$ is
 \begin{equation}
    -\frac{\Delta{r_2}(t-1)}{b_2}<\Delta{r_1}(t-1)<-{b_1}\Delta{r_2}(t-1).
     \label{Eq:A4}
\end{equation}
Since $-{\Delta{r_2}(t-1)}/{b_2}>-{b_1}\Delta{r_2}(t-1)$ when $\Delta{r_2}(t-1)>0$, Eq. (\ref{Eq:A4}) cannot be satisfied, therefore this case is impossible.

For the case when $\Delta{r_1^e}(t)>0$ and $\Delta{r_2^e}(t)>0$ are satisfied, according to Eq. (\ref{Eq:13}), the condition will be
\begin{equation}
    -{b_1}\Delta{r_2}(t-1)<\Delta{r_1}(t-1)<-\frac{\Delta{r_2}(t-1)}{b_2}.
     \label{Eq:A5}
\end{equation}
Under this condition, the expected returns are positively correlated, and its probability becomes smaller when $b_j$ $(j=1, 2)$ is closer to -1.

We next discuss the other two cases when the expected returns are negatively correlated.
If $\Delta{r_1^e}(t)>0$ and $\Delta{r_2^e}(t)<0$ are satisfied, according to Eq. (\ref{Eq:13}), the condition will be
\begin{equation}
          \Delta{r_1}(t-1)>-\frac{\Delta{r_2}(t-1)}{b_2}.
          \label{Eq:A6}
\end{equation}
When $b_2$ approaches -1, the lower limit value of $\Delta{r_1}(t-1)$ approaches $\Delta{r_2}(t-1)$. Therefore, the value range of $\Delta{r_1}(t-1)$ becomes larger, and the probability of satisfying $\Delta{r_1^e}(t)>0$ and $\Delta{r_2^e}(t)<0$ is larger. The negative correlation of the expected returns is stronger.
Similarly, if $\Delta{r_1^e}(t)<0$ and $\Delta{r_2^e}(t)>0$ are satisfied, the negative correlation of the expected returns is stronger when $b_1$ approaches -1.

For two of the three valid cases, the expected returns are negatively correlated with large probability. The correlation are negative on average, and the negative correlation becomes stronger when $b_j$ $(j=1, 2)$ approaches -1.

(b) If $\Delta{r_1}(t-1)<0$ and $\Delta{r_2}(t-1)<0$, the analysis is similar to Appendix A \uppercase\expandafter{\romannumeral2} (a). Here, the negative correlation of the expected returns becomes stronger when $b_j$ $(j=1, 2)$ approaches -1.

(c) If $\Delta{r_1}(t-1)>0$ and $\Delta{r_2}(t-1)<0$, then
\begin{equation}
\left\{
             \begin{array}{lr}
             \Delta{r_1^e}(t)=\Delta{r_1}(t-1)+{b_1}\Delta{r_2}(t-1)>0 &  \\
             \Delta{r_2^e}(t)=\Delta{r_2}(t-1)+{b_2}\Delta{r_1}(t-1)<0. &
             \end{array}
\right.
\label{Eq:A7}
\end{equation}
Here, the expected return of stock 1 increases while the expected return of stock 2 decreases, and the expected return of the stocks move in opposite directions. Therefore the correlation of the expected returns is negative.

(d) If $\Delta{r_1}(t-1)<0$ and $\Delta{r_2}(t-1)>0$, the analysis is similar to Appendix A \uppercase\expandafter{\romannumeral2} (c). Here, the expected return of stock 1 decreases while the expected return of stock 2 increases, and the expected return of the stocks move in opposite directions. Hence the correlation of the expected returns is negative.

\uppercase\expandafter{\romannumeral3}. $0<b_1<1$ and $-1<b_2<0$

(b) If $\Delta{r_1}(t-1)<0$ and $\Delta{r_2}(t-1)<0$, $\Delta{r_1^e}(t)$ is always negative. There are two possible cases for $\Delta{r_2^e}(t)$, namely $\Delta{r_2^e}(t)<0$ and $\Delta{r_2^e}(t)>0$.

If $\Delta{r_1^e}(t)<0$ and $\Delta{r_2^e}(t)<0$ are satisfied, according to Eq. (\ref{Eq:13}), the condition will be
\begin{equation}
          \Delta{r_2}(t-1)<-{b_2}\Delta{r_1}(t-1).
          \label{Eq:A8}
\end{equation}
When $b_2$ approaches -1, the upper limit value of $\Delta{r_2}(t-1)$ approaches $\Delta{r_1}(t-1)$. Therefore, the value range of $\Delta{r_2}(t-1)$ becomes smaller, the probability of satisfying $\Delta{r_1^e}(t)<0$ and $\Delta{r_2^e}(t)<0$ becomes smaller. The probability that the expected returns of both stocks decrease simultaneously becomes smaller, i.e., the positive correlation of the expected returns becomes weaker.

Similarly, if $\Delta{r_1^e}(t)<0$ and $\Delta{r_2^e}(t)>0$ are satisfied, the negative correlation of the expected returns becomes stronger when $b_2$ approaches -1.

(d) If $\Delta{r_1}(t-1)<0$ and $\Delta{r_2}(t-1)>0$, $\Delta{r_2^e}(t)$ is always positive. There are two possible cases for $\Delta{r_1^e}(t)$, namely $\Delta{r_1^e}(t)<0$ and $ \Delta{r_1^e}(t)>0$.

If $\Delta{r_1^e}(t)>0$ and $\Delta{r_2^e}(t)>0$ are satisfied, according to Eq. (\ref{Eq:13}), the condition will be

\begin{equation}
           \Delta{r_2}(t-1)>-\frac{\Delta{r_1}(t-1)}{b_1}.
           \label{Eq:A9}
\end{equation}
When $b_1$ approaches 1, the upper limit value of $\Delta{r_2}(t-1)$ approaches $-\Delta{r_1}(t-1)$. Therefore, the value range of $\Delta{r_2}(t-1)$ becomes larger, the probability of satisfying $\Delta{r_1^e}(t)>0$ and $\Delta{r_2^e}(t)>0$ becomes larger. The probability that the expected returns of both stocks increase simultaneously becomes larger, i.e., the positive correlation of the expected returns becomes stronger.

Similarly, if $\Delta{r_1^e}(t)<0$ and $\Delta{r_2^e}(t)>0$ are satisfied, the negative correlation of the expected returns becomes weaker when $b_1$ approaches 1.

\uppercase\expandafter{\romannumeral4}. $-1<b_1<0$ and $0<b_2<1$

(a) If $\Delta{r_1}(t-1)>0$ and $\Delta{r_2}(t-1)>0$, $\Delta{r_2^e}(t)$ is always positive. There are two possible cases for $\Delta{r_1^e}(t)$, namely $\Delta{r_1^e}(t)>0$ and $\Delta{r_1^e}(t)<0$.

If $\Delta{r_1^e}(t)>0$ and $\Delta{r_2^e}(t)>0$ are satisfied, according to Eq. (\ref{Eq:13}), the condition will be
\begin{equation}
          \Delta{r_1}(t-1)>-{b_1}\Delta{r_2}(t-1).
          \label{Eq:A10}
\end{equation}
When $b_1$ approaches -1, the lower limit value of $\Delta{r_1}(t-1)$ approaches $\Delta{r_2}(t-1)$. Therefore, the value range of $\Delta{r_1}(t-1)$ becomes smaller, the probability of satisfying $\Delta{r_1^e}(t)>0$ and $\Delta{r_2^e}(t)>0$ becomes smaller. The probability that the expected returns of both stocks increase simultaneously becomes smaller, i.e., the positive correlation of the expected returns becomes weaker.

Similarly, if $\Delta{r_1^e}(t)<0$ and $\Delta{r_2^e}(t)>0$ are satisfied, the negative correlation of the expected returns is stronger when $b_1$ approaches -1.

(b) If $\Delta{r_1}(t-1)<0$ and $\Delta{r_2}(t-1)<0$, the analysis here is similar to Appendix A \uppercase\expandafter{\romannumeral4} (a). Here, the positive (negative) correlation of the expected returns becomes weaker (stronger) when $b_1$ approaches -1.

(c) If $\Delta{r_1}(t-1)>0$ and $\Delta{r_2}(t-1)<0$, $\Delta{r_1^e}(t)$ is always positive. There are two possible cases for $\Delta{r_2^e}(t)$, namely $\Delta{r_2^e}(t)>0$ and $\Delta{r_2^e}(t)<0$.

If $\Delta{r_1^e}(t)>0$ and $\Delta{r_2^e}(t)>0$ are satisfied, according to Eq. (\ref{Eq:13}), the condition will be
\begin{equation}
           \Delta{r_1}(t-1)>-\frac{\Delta{r_2}(t-1)}{b_2}.
           \label{Eq:A11}
\end{equation}
When $b_2$ approaches 1, the lower limit value of $\Delta{r_1}(t-1)$ approaches $-\Delta{r_2}(t-1)$. Therefore, the value range of $\Delta{r_1}(t-1)$ becomes larger, the probability of satisfying $\Delta{r_1^e}(t)>0$ and $\Delta{r_2^e}(t)>0$ is larger. The positive correlation of the expected returns is stronger.

Similarly, if $\Delta{r_1^e}(t)>0$ and $\Delta{r_2^e}(t)<0$ are satisfied, the negative correlation of the expected returns is weaker when $b_2$ approaches 1.

(d) If $\Delta{r_1}(t-1)<0$ and $\Delta{r_2}(t-1)>0$, the analysis is similar to Appendix A \uppercase\expandafter{\romannumeral4} (c). Here, the positive (negative) correlation of the expected returns becomes stronger (weaker) when $b_2$ approaches 1.


\bibliographystyle{elsarticle-harv}
\bibliography{E:/Papers/Auxiliary/Bibliography}







\end{document}